\documentclass[12pt]{elsarticle}
\usepackage{amsmath,amssymb,amsfonts}
\usepackage{multirow}
\usepackage{subfig}
\usepackage{soul}
\usepackage{color}
\usepackage{bm}
\usepackage{hyperref}
\usepackage{rotating}
\usepackage{graphicx}
\usepackage{amssymb}
\usepackage{lineno}
\biboptions{numbers,sort&compress}
\journal{arXiv}%

\newcommand{\SSIM}{\ensuremath{\operatorname{SSIM}}}
\newcommand{\pSNR}{\ensuremath{\operatorname{pSNR}}}

\begin{document}
\tolerance=999
\sloppy

\begin{frontmatter}


\title{Multi-Coil MRI Reconstruction Challenge - Assessing Brain MRI Reconstruction Models and their Generalizability to Varying Coil Configurations}

\author[label2,hbi,label1]{Youssef Beauferris}
\author[label9,label10,labeln]{Jonas Teuwen}
\author[label11]{Dimitrios Karkalousos}
\author[label9,label10]{Nikita Moriakov}
\author[label11]{Matthan Caan}
\author[label10,labeln]{George Yiasemis} 
\author[feec]{L\'{i}via Rodrigues}
\author[ic]{Alexandre Lopes}
\author[ic]{Helio Pedrini}
\author[feec]{Let\'{i}cia Rittner}
\author[label6]{Maik Dannecker}
\author[label6]{Viktor Studenyak}
\author[label6]{Fabian Gr\"{o}ger}
\author[label6]{Devendra Vyas}
\author[label6]{Shahrooz Faghih-Roohi}
\author[label7]{Amrit Kumar Jethi}
\author[label7]{Jaya Chandra Raju}
\author[label7,label8]{Mohanasankar Sivaprakasam}
\author[label3,hbi]{Mike Lasby} 
\author[mcmaster01,mcmaster02]{Nikita Nogovitsyn} 
\author[label2,hbi,label1]{Wallace Loos}
\author[label2,hbi,label1]{Richard Frayne}
\author[hbi,label3]{Roberto Souza}

\address[label2]{Radiology and Clinical Neurosciences, University of Calgary, Canada}
\address[hbi]{Hotchkiss Brain Institute, University of Calgary, Canada}
\address[label1]{Seaman Family MR Research Centre, Foothills Medical Center, Canada}
\address[label9]{Department of Medical Imaging, Radboud University Medical Center, the Netherlands}
\address[label10]{Department of Radiation Oncology, Netherlands Cancer Institute, the Netherlands}
\address[labeln]{ICAI-AI for Oncology, University of Amsterdam,   the Netherlands}
\address[label11]{Department of Biomedical Engineering and Physics, Amsterdam UMC, University of Amsterdam, the Netherlands}
\address[feec]{School of Electrical and Computer Engineering, University of Campinas, Brazil}
\address[ic]{Institute of Computing, University of Campinas, Brazil}
\address[label6]{Computer Aided Medical Procedures, Technical University of Munich, Germany}
\address[label7]{Indian Institute of Technology Madras (IITM), India}
\address[label8]{Healthcare Technology Innovation Centre (HTIC), IITM,  India}

\address[label3]{Department of Electrical and Software Engineering, University of Calgary, Canada}
\address[mcmaster01]{Centre for Depression and Suicide Studies, St. Michael’s Hospital, Canada}
\address[mcmaster02]{Mood Disorders Program, Department of Psychiatry and Behavioural Neurosciences, McMaster University, Canada}

\begin{abstract}
Deep-learning-based brain magnetic resonance imaging (MRI) reconstruction methods have the potential to accelerate the MRI acquisition process. Nevertheless, the scientific community lacks appropriate benchmarks to assess MRI reconstruction quality of high-resolution brain images, and  evaluate how these proposed algorithms will behave in the presence of small, but expected data distribution shifts. The Multi-Coil Magnetic Resonance Image (MC-MRI) Reconstruction Challenge provides a benchmark that aims at addressing these issues, using a large dataset of high-resolution, three-dimensional, T1-weighted  MRI scans. The challenge has two primary goals: 1) to compare different MRI reconstruction models on this dataset and 2) to assess the generalizability of these models to data acquired with a different number of receiver coils. In this paper, we describe the challenge experimental design, and summarize the results of a set of baseline and state of the art brain MRI reconstruction models. We provide relevant comparative information on the current MRI reconstruction state-of-the-art  and highlight the challenges of obtaining generalizable models that are required prior to broader clinical adoption. The MC-MRI benchmark data, evaluation code and current challenge leaderboard are publicly available. They provide an objective performance assessment for future developments in the field of brain MRI reconstruction.  
\end{abstract}


\end{frontmatter}


\section{Introduction}
\noindent Brain magnetic resonance imaging (MRI) is a commonly used diagnostic imaging modality. It is a non-invasive technique that provides images with excellent soft-tissue contrast. Brain MRI produces a wealth of information, which often leads to definitive diagnosis of a number of neurological conditions, such as cancer and stroke. Furthermore, it is broadly adopted in neuroscience and other research domains. MRI data acquisition occurs in the Fourier or spatial-frequency domain, more commonly referred to as $k$-space. Image reconstruction consists of transforming the acquired $k$-space raw data into interpretable images. Traditionally, data is collected following the Nyquist sampling theorem~\cite{RN247}, and for a single-coil acquisition, a simple inverse Fourier transform operation is often sufficient to reconstruct an image. However, the fundamental physics, practical engineering aspects and biological tissue response factors underlying the MRI data acquisition process, makes fully sampled acquisitions inherently slow. These limitations represent a crucial drawback when MRI is compared to other medical imaging modalities and impacts both patient tolerance of the procedure and throughput, as well as more broadly neuroimaging research.

Parallel imaging (PI)~\cite{sense, grappa,deshmane2012parallel}
and compressed sensing (CS)~\cite{lustig2007,liang2009accelerating} are two proven approaches that are able to reconstruct high-fidelity images from sub-Nyquist sampled acquisitions. PI techniques leverage the spatial information available across multiple, spatially distinct, receiver coils to allow reconstruction of undersampled $k$-space data. Techniques, such as generalized autocalibrating partially parallel acquisition (GRAPPA)~\cite{grappa}, which operates in the $k$-space domain, and sensitivity encoding for fast MRI (SENSE)~\cite{sense}, which works in the image domain, are currently used clinically. CS methods leverage image sparsity properties to improve reconstruction quality from undersampled $k$-space data. Some CS techniques, such as compressed SENSE~\cite{liang2009accelerating}, have also seen clinical adoption. Those PI and CS methods that have been approved for routine clinical use are generally restricted to relatively conservative acceleration factors (\textit{e.g.}, $ R = 2\times$ to $3\times$ acceleration). Currently employed comprehensive brain MRI scanning protocols, even those that use PI and CS, typically require between 30 and 45 minutes per patient procedure. Longer procedural times increase patient discomfort, thus lessening the likelihood of patient acceptance. It also increases susceptibility to both voluntary and involuntary motion artifacts. 

In 2016, the first deep-learning-based MRI reconstruction models were presented~\cite{wang2016,sun2016}. The excellent initial results obtained by these models caught the attention of the MR imaging community, and  subsequently, dozens of deep-learning-based MRI reconstruction models were proposed, \textit{cf.},~\cite{hammernik2018, sun2016, wang2016, schlemper2017, dedmari2018complex,Schlemper2018StochasticDC, kwon2017parallel, RN323, dautomap, pawar2019deep, eo2018kiki, qin2019convolutional, mardani2019deep, zhang2018multi, semantic_interpretability, akccakaya2019scan, souza2018hybrid, RN305, RN307, RN289, RN311, zengK2019, sriram2020grappanet, zhou2020dudornet,hosseini2020dense} provides a partial listing. Many of these studies demonstrated superior quantitative results from deep-learning-based methods compared to non-deep-learning-based MRI reconstruction algorithms~\cite{hammernik2018,schlemper2017,knoll2020advancing}. These new methods are also capable of accelerating MRI examinations beyond traditional PI and CS methods. There is good evidence that deep-learning-based MRI reconstruction methods can accelerate MRI examinations by factors greater than 10~\cite{zbontar2018fastmri,souza2020enhanced}.  

A significant drawback, that hinders the progress of the brain MRI reconstruction field, is the lack of benchmark datasets. Importantly, the lack of benchmarks makes comparison of different methods challenging. The fastMRI effort~\cite{zbontar2018fastmri} is an important initiative that provides large volumes of raw MRI $k$-space data. The initial release of the fastMRI dataset provided two-dimensional (2D) MR acquisitions of the knee. A subsequent release added 2D brain MRI data with $5$ mm slice thickness, which was used for the 2020 \textit{fastMRI} challenge \cite{muckley2021results}. The \textit{Calgary-Campinas}~\cite{souza2017} initiative contains numerous sets of brain imaging data. For the purposes of this benchmark, we expanded the \textit{Calgary-Campinas} initiative to include MRI raw data from three-dimensional (3D), high-resolution acquisitions. High-resolution images are crucial for many neuroimaging applications. Also importantl, 3D acquisitions allow for undersampling along two phase encoding dimensions, instead of one for 2D imaging. This potentially allows for further MRI acceleration. These $k$-space datasets correspond to either 12- or 32-channel data.

The goals of the Multi-Coil Magnetic Resonance Image (MC-MRI - \url{https://www.ccdataset.com/mr-reconstruction-challenge})) Reconstruction   Challenge are to provide benchmarks that help improve the quality of brain MRI reconstruction, facilitate comparison of different reconstruction models, better understand the difficulties related to clinical adoption of these models, and investigate the upper limits of MR acceleration. The specific objectives of the challenge are:
\begin{enumerate}
    \item Compare the performance of different brain MRI reconstruction models on a large dataset, and
    \item Assess the generalizability of these models to datasets acquired with different coils.
\end{enumerate}

\noindent The results presented in this report correspond to benchmark submissions received up to November 20$^{th}$, 2021. Four baseline solutions and three new benchmark solutions were presented and discussed during an online session at the Medical Imaging Deep Learning Conference held on July 9$^{th}$, 2020\footnote{See video of session at https://www.ccdataset.com/mr-reconstruction-challenge/mc-mrrec-2020-midl-recording}. Two additional benchmark solutions were submitted after the online session. Collectively, these results provide a relevant performance summary of some state of the art MRI reconstruction approaches, including different model architectures, processing strategies, and emerging metrics for training and assessing reconstruction models. The MC-MRI Reconstruction Challenge is ongoing and open to new benchmark submissions\footnote{See current leaders for the individual challenge tracks at https://www.ccdataset.com/}. A public code repository with instructions on how to load the data, extract the benchmark metrics, and baseline reconstruction models is available at \url{https://github.com/rmsouza01/MC-MRI-Rec}.

\section{Materials and Methods}

\subsection{Calgary-Campinas Raw MRI Dataset}

The data used in this challenge were acquired as part of the Calgary Normative Study~\cite{mccreary2020calgary}, which is a multi-year, longitudinal project that investigates normal human brain ageing by acquiring quantitative MRI data using a protocol approved by our local research ethics board.
Raw data from T1-weighted volumetric imaging were acquired, anonymized and incorporated into the \textit{Calgary-Campinas} (\textit{CC}) dataset~\cite{souza2017}. The publicly accessible dataset currently provides $k$-space data from 167 3D, T1-weighted, gradient-recalled echo, 1 mm$^3$ isotropic sagittal acquisitions collected on a clinical 3-T MRI scanner (Discovery MR750; General Electric Healthcare, Waukesha, WI). The brain scans are from presumed healthy subjects (mean $\pm$ standard deviation age: $44.5 \pm 15.5$ years;  range: $20$ years to $80$ years; $71/167$ ($42.5\%$) male).

The datasets were acquired using either a 12-channel (117 scans, $70.0\%$) or 32-channel receiver coil (50 scans, $30.0\%$). Acquisition parameters were TR/TE/TI = 6.3 ms / 2.6 ms / 650 ms (93 scans, $55.7\%$) or TR/TE/TI = 7.4 ms / 3.1 ms / 400 ms (74 scans, $44.3\%$), with 170 to 180 contiguous 1.0 mm slices and a field of view of $256$ mm $\times$ $218$ mm. The acquisition matrix size $[N_x, N_y, N_z]$ for each channel was $[256, 218, 170-180]$, where $x$, $y$, and $z$ denote readout, phase-encode and slice-encode directions, respectively. In the slice-encode ($k_z$) direction, only 85\% of the $k$-space data were collected; the remainder (15\% of 170-180) was zero-filled. This partial acquisition technique is common practice in MRI. The average scan duration is 341 seconds.  Because $k$-space undersampling only occurs in the phase-encode and slice-encode directions, the 1D inverse Fourier transform (iFT) along $k_x$ was automatically performed by the scanner and hybrid $(x, k_y, k_z)$ datasets were provided. This pre-processing effectively allows the MRI reconstruction problem to be treated as a 2D problem (in $k_y$ and $k_z$). The partial Fourier reference data was reconstructed by taking the 2D iFT along the $k_y-k_z$ plane for each individual channel and combining these using the conventional square-root sum-of-squares algorithm~\cite{larsson2003snr}.

\subsection{MC-MRI Reconstruction Challenge Description}
The MC-MRI Reconstruction Challenge was designed to be an ongoing investigation that will be disseminated through  a combination of in-person sessions at meetings and virtual sessions, supplemented by periodic on-line submissions and updates. The benchmark is readily extensible and more data, metrics and research questions are expected to be added in further updates. Individual research groups are permitted to make multiple submissions. The processing of submissions is semi-automated and it takes on average 48 hours to generate an update of the benchmark leaderboard.

Currently, the MC-MRI Reconstruction Challenge is split into two separate tracks. Teams can decide whether to submit a solution to just one track or to both tracks. Each track has a separate leaderboard. The tracks are:

\begin{itemize}
    \item \textbf{Track 01:} Teams had access to 12-channel data to train and validate their models. Models submitted are evaluated only using the 12-channel test data. 
    \item \textbf{Track 02:} Teams had access to 12-channel data to train and validate their models. Models submitted are evaluated for both the 12-channel and 32-channel test data
\end{itemize}

\noindent In both tracks, the goal is to assess the brain MR image reconstruction quality and in particular note any loss of high-frequency details, especially at the higher acceleration rates. By having two separate tracks, we hoped to determine whether a generic reconstruction model trained on data from one coil would have decreased performance when applied to data from another coil.

Two MRI acceleration factors were tested: $R = 5$ and $R = 10$. These factors were chosen intentionally to exceed the acceleration factors typically used clinically with PI and CS methods. A Poisson disc distribution sampling scheme, where the center of $k$-space was fully sampled within a circle of radius of 16 pixels to preserve the low-frequency phase information, was used to achieve these acceleration factors. For brevity, we have only reported the results for $R=5$, but the online challenge leaderboard contains the results for both acceleration factors.

The training, validation and test split of the challenge data is summarized in Table~\ref{dataset_summary}. The initial 50 and last 50 slices in each participant image volume were removed because they have little anatomy present. The fully sampled $k$-space data of the training and validation sets were made public for teams to develop their models. Pre-undersampled  $k$-space data corresponding to the test sets were provided for the teams for accelerations of $R=5$ and $R=10$.

\begin{table}[!h]
\caption{Summary of the raw MRI $k$-space datasets used in the first edition of the challenge. Reported are the number of slices in the test sets after removal of the initial 50 and last 50 slices (see text).}
\label{dataset_summary}
\centering
\begin{tabular}{c|c|c|c}
\hline
\textbf{Coil} &  \textbf{Category} &  \textbf{\# of datasets} &  \textbf{\# of slices}  \\ \hline
\multirow{3}{*}{12-channel}  & Train & 47 & 12,032 \\  \cline{2-4}
& Validation & 20 & 5,120  \\\cline{2-4}
& Test & 50 & $7,800$  \\ \hline
32-channel & Test & 50 & $7,800$ \\ \hline
\end{tabular}
\end{table}

\subsection{Quantitative Metrics}
In order to measure the quality of the image reconstructions, three commonly used, quantitative performance metrics were selected: peak signal-to-noise ratio (pSNR), structural similarity (SSIM) index~\cite{wang2004}, and visual information fidelity (VIF)~\cite{sheikh2006}. The choice of performance metrics is challenging and it is recognized that objective measures such as pSNR, SSIM and VIF may not correlate well with subjective human image quality assessments. Nonetheless, these metrics provide a broad basis to assess model performance in this challenge.

The pSNR is a metric commonly used for MRI reconstruction assessment and consists of the log ratio between the maximum value of the reference reconstruction and the root mean squared error (RMSE):

\begin{equation}
    \pSNR(y, \hat{y}) = 20\log_{10}\left(\frac{\max(y)}{{\operatorname{RMSE}}}\right) = 20\log_{10}\left(\frac{\max(y)}{\sqrt{\frac{1}{M}\sum_{i=1}^M[y(i) - \hat{y}(i)]^2}}\right),
    \label{eq:ssim}
\end{equation}

\noindent where $y$ is the reference image, $\hat{y}$ is the reconstructed image, and $M$ is the number of pixels in the image. Higher pSNR values represent higher-fidelity image reconstructions. However, pSNR does not take into consideration factors involved in human vision. For this reason, increased pSNR can suggest that reconstructions are of higher quality, when in fact they may not be as well perceived by the human visual system. 

Unlike pSNR, SSIM and VIF are metrics that attempt to model aspects of the human visual system. SSIM considers biological factors such as luminance, contrast and structural information. SSIM is computed using:
\begin{equation}
    \SSIM(x, \hat{x}) = \frac{(2\mu_{x}\mu_{\hat{x}}+c_1)(2\sigma_{x\hat{x}}+c_2)}{(\mu{^2}_{x}+\mu{^2}_{\hat{x}} + c_1)(\sigma{^2}_{x}+\sigma{^2}_{\hat{x}} + c_2)}
    \label{eq:ssim}
\end{equation}
\noindent where $x$ and $\hat{x}$ represent corresponding image windows from the reference image and the reconstructed image, respectively; $\mu_x$ and $\sigma_x$ represent the mean and standard deviation inside the image window, $x$; and $\mu_{\hat{x}}$ and $\sigma_{\hat{x}}$ represent the mean and standard deviation inside the reconstructed image window, $\hat{x}$. The constants $c_1$ and $c_2$ are used to avoid numerical instability. SSIM values for non-negative images are within $[0,1]$, where 1 represents two identical images.

The VIF metric is based on natural scene statistics~\cite{simoncelli2001, wilson2008}.  
VIF models the natural scene statistics based on a Gaussian scale mixture model in the wavelet domain, and additive white Gaussian noise is used to model the human visual system. The natural scene of the reference image is modeled into wavelet components (C) and the human visual system is modeled by adding zero-mean white Gaussian noise in the wavelet domain (N), which results in the perceived reference image (E = C + N). In the same way, the reconstructed image, which is called the distorted image, is also modeled by a natural scene model (D) and the human visual system  model (N'), leading to the perceived distorted image (F = D + N').  The VIF is given by the ratio of the mutual information of I(C, F) and I(C, E):
 \begin{equation}
 \text{VIF} = \frac{I(C,F)}{I(C,E)},
 \end{equation}
 where $I$ represents the mutual information. 
 
 Mason \textit{et al.}~\cite{mason2019comparison} investigated the VIF metric for assessing MRI reconstruction quality. Their results indicated that it has a stronger correlation with subjective radiologist opinion about MRI quality than other metrics such as pSNR and SSIM. The VIF Gaussian noise variance was set to $0.4$ as recommended in~\cite{mason2019comparison}. All metrics were computed slice-by-slice in the test set. The reference and reconstructed images were normalized by dividing them by the maximum value of the reference image.

\subsection{Visual Assessment}
An expert observer (NN) with over five years of experience analyzing brain MR images and manually segmenting complex structures, such as the hippocampus and hypothalamus, visually inspected 25 randomly selected volumes for the 12-channel test set and other 25 volumes for the 32-channel test set for the best two submissions  as determined from the quantitative metrics. The best two submissions were obtained by sorting the weighted average ranking. The weighted average ranking was generated by applying pre-determined weights to the ranking of the three individual quantitative metrics (0.4 for VIF, 0.4 for SSIM and 0.2 for pSNR). We chose to give higher weights to VIF and SSIM because they have a better correlation with the human perception of image quality.  

The visual assessment of the images was done by comparing the machine-learning-based reconstructions to the fully sampled reference images. This allowed the observer to distinguish between data acquisition related  quality issues (\textit{e.g.}, motion) and problems associated with image reconstruction. The image quality assessment focused mostly on overall image quality and how well defined was the contrast between white-matter, gray-matter, and other relevant brain structures. The goal of the visual assessment was to compare the quality of the reconstructed MR images against the fully sampled reference images and not to compare the quality of the different submissions, because the benchmark is ongoing and we wanted to account for potential observer memory bias effects~\cite{kalm2018visual} in the qualitative metrics due to the difference between submission dates of the different solutions to the benchmark (\textit{i.e.}, future submissions will be visually assessed at different dates compared to current submissions).  

\subsection{Models}
Track 01 of the challenge included four baseline models, selected from the literature. These models are the zero-filled reconstruction, the U-Net model~\cite{RN254}, the WW-net model~\cite{souza2020dual}, and the hybrid-cascade model~\cite{souza19a}. To date, Track 01 has received six independent submissions from  ResoNNance~\cite{DIRECTTOOLKIT} (two different models), The Enchanted (two different models), TUMRI, and M-L UNICAMP teams. 

The ResoNNance 1.0 model submission was a recurrent inference machine~\cite{lonning2019recurrent}, ResoNNance 2.0 was a recurrent variational network~\cite{yiasemis2021}. The Enchanted 1.0 model was inspired by~\cite{lee2018deep}, where they used  magnitude and phase networks, followed by a VS-net architecture~\cite{duan2019vs}. The Enchanted 2.0 used an end-to-end variational network~\cite{sriram2020end}, and it was the only submission that used self-supervised learning~\cite{chen2019self} to initialize  their model. The pretext task to initialize their models was the prediction of image rotations~\cite{rotation}. TUMRI used a similar model to the WW-net, but they implemented complex-valued operations~\cite{deepcomplexnet}. They used a linear combination of VIF and multi-scale SSIM~\cite{msssim} as their loss function. M-L UNICAMP used a hybrid model with parallel network branches operating in $k$-space and image domains. Links to the source code for the different models are available in the benchmark repository. Some of the Track 01 models were designed to work with a specific number of coil channels, thus they were not submitted to Track 02 of the challenge.

Track 02 of the challenge included two baseline models (zero-filled reconstruction and the U-Net model). ResoNNance and The Enchanted teams submitted two models each to Track 02. The models submitted by ResoNNance and The Enchanted teams were the same models that were used for Track 01 of the challenge. Table~\ref{summary_models} summarizes the  processing domains (image, $k$-space or dual/hybrid), the presence of elements, such as coil sensitivity estimation, data consistency, and the loss function used during training of the models. For more details about the models, we refer the reader to the source publications or the code repositories for the unpublished work.

\begin{table}[!ht]
\caption{Summary of the submissions including processing domain, presence of coil sensitivity estimation (SE), presence of data consistency (DC), and basis of the training loss functions. $\ast$ indicates a baseline model. Loss functions: Mean Absolute Error (MAE), Structural Similarity (SSIM), Mean Squared Error (MSE), Multi-Scale SSIM (MS-SSIM), and Visual Information Fidelity (VIF). }
\label{summary_models}
\centering
\resizebox{\columnwidth}{!}{%
\begin{tabular}{c|c|c|c|c|c}
\hline
\textbf{Model} &  \textbf{Domain} &  \textbf{SE} &  \textbf{DC} &  \textbf{Loss function} \\ \hline
\textbf{ResoNNance 2.0} &  Hybrid &  Yes & Yes &  MAE and SSIM\\ \hline
\textbf{The Enchanted 2.0} &  Image & Yes & Yes & cross entropy (pretext) and SSIM (main task) \\ \hline
\textbf{ResoNNance 1.0} &   Image & Yes & Yes & MAE and SSIM \\ \hline
\textbf{The-Enchanted 1.0} &   Image & Yes & Yes & MSE (first step) and SSIM (second step)\\ \hline
\textbf{TUMRI} & Hybrid & No & Yes & MS-SSIM and VIF  \\ \hline
\textbf{WW-Net$\ast$} &  Hybrid & No & Yes & MSE \\ \hline
\textbf{Hybrid-cascade$\ast$} &  Hybrid & No & Yes & MSE \\ \hline
\textbf{M-L UNICAMP} &  Hybrid & No & Yes & MSE \\ \hline
\textbf{U-Net$\ast$} &  Image & No & No & MSE \\ \hline
\textbf{Zero-filled$\ast$} &  N/A  & No & N/A & N/A \\ \hline
\end{tabular}}
\end{table}

\section{Results}


\subsection{Track 01}
The quantitative results for Track 01 are summarized in Table~\ref{t1_table}. There were in total ten models (four baseline and six submitted) in Track 01. The zero-filled and U-Net reconstructions had the worst results. The M-L UNICAMP, Hybrid Cascade, WW-net, and TUMRI models were next with similar results in terms of SSIM and pSNR. Notably, the TUMRI submission achieved the second highest VIF metric. ResoNNance and The Enchanted teams' submissions achieved the highest overall scores on the quantitative metrics. The ResoNNance 2.0 submission had the best SSIM and pSNR metrics and the fourth best VIF metric. The Enchanted 1.0 submission obtained the best VIF metric. The Enchanted 2.0 submission achieved the second best SSIM metric, and the third best VIF and pSNR metrics. Representative reconstructions resulting from the different models for $R=5$ are shown in Figure~\ref{sample_rec_12_channel}. 


Twenty five images in the test set were visually assessed by our expert observer for the two best submisisons (ResoNNance 2.0 and The Enchanted 2.0). Out of the 50 images assessed by the expert observer, only two ($4.0\%$) were deemed to have minor deviations from common anatomical borders. Twenty seven images ($54.0\%$) were deemed to have similar quality to the fully sampled reference, and 21 ($42.0\%$) were rated as having similar quality when compared to the reference, but exhibited filtering of the noise in the image background.

\subsection{Track 02}
Two teams, ResoNNance and The Enchanted, submitted a total of four models to Track 02 of the benchmark. Their results were compared to two baseline techniques. The models submitted to Track 02, except for the U-Net baseline, which has a higher input dimension (\textit{i.e.}, the input dimensions depends on the number of receiver coils), were the same as the models submitted for Track 01, so for the 12-channel test dataset, the results are the same as in Track 01 (see Table~\ref{t1_table}). 

The results for Track 02 using the 32-channel test set are summarized in Table~\ref{t2_table_32}. For the 32-channel test dataset, The Enchanted 2.0 submission obtained the best VIF and pSNR metrics, and second best SSIM score. The ResoNNance 2.0 submission obtained the best SSIM metric, second best pSNR, and third best VIF metrics. The ResoNNance 1.0 submission obtained the third best SSIM and pSNR metrics, and second best VIF. The Enchanted 1.0 submission obtained the fourth best SSIM and VIF, and fifth best pSNR.  The zero-filled and U-Net reconstructions obtained the worse results. Representative reconstructions resulting from the different models are depicted in Figure~\ref{sample_rec_32_channel}. 

Twenty five images in the test set were visually assessed by our expert observer for the two best submisisons (ResoNNance 2.0 and The Enchanted 2.0). Out of the 50 images assessed by the expert observer, 14 ($28.0\%$) were deemed to have  deviations from common anatomical borders. Thirty four images ($68.0\%$) were deemed to have similar quality to the fully sampled reference, and only two images ($4.0\%$) were rated as having similar quality when compared to the reference, but exhibited filtering of the noise in the image background.

\section{Discussion}
The first track of the challenge compared ten different reconstruction models (Table~\ref{t1_table}). As expected, the zero-filled reconstruction, which does not involve any training from the data, universally had the poorest results. The second worst technique was the U-Net model, which used as input the channel-wise zero-filled reconstruction and tried to recover the high-fidelity image. The employed U-Net~\cite{RN254} model did not include any data consistency steps. The remaining eight models all includes a data consistency step, which seems to be an essential step for high-fidelity image reconstruction, as has been previously highlighted in~\cite{schlemper2017,eo2018kiki}.

The M-L UNICAMP model explored parallel architectures that operated both in the $k$-space and image domains. M-L UNICAMP had the eighth lowest pSNR and VIF metrics, and the seventh lowest SSIM score. In contrast, the top ranked methods were either cascaded networks (Hybrid-cascade, WW-net, TUMRI, The Enchanted 1.0 and 2.0) or recurrent methods (ResoNNance 1.0 and 2.0).

The top four models in the benchmark were the ResoNNance 1.0 and 2.0 and The Enchanted 1.0 and 2.0 submissions. These four models estimated coil sensitivities and combined the coil channels, which made these models flexible and capable of working with datasets acquired with an arbitrary number of receiver coils. With the exception of ResoNNance 2.0, which is a hybrid model, these are image-domain methods. The other better performing models (M-L UNICAMP, Hybrid Cascade, WW-net, and TUMRI) used an approach that receives all coil channels as input, making these models tailored to a specific coil configuration (\textit{i.e.}, number of channels). Though the methods that combined the channels before reconstruction, such as from ResoNNance and The Enchanted teams, demonstrated the best results so far, it is still unclear if this approach is superior to models that do not combine the channels before reconstruction. A recent work~\cite{sriram2020grappanet} indicated that the separate channel approach may be advantageous compared to models that combine the $k$-space channels before reconstruction.

\begin{figure*}[!ht]
\begin{centering}
\includegraphics[width=0.95\textwidth]{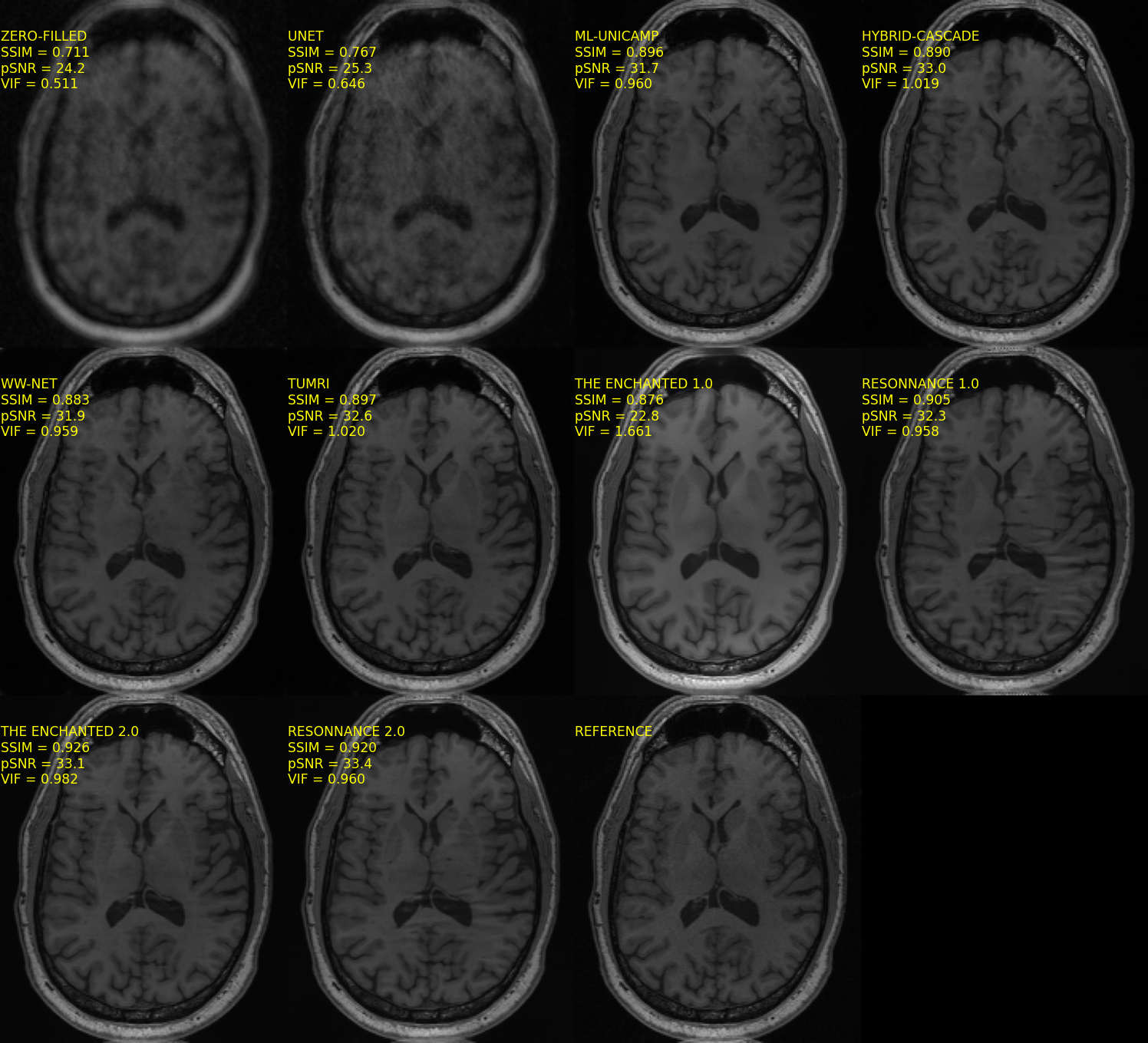}\\
\caption{Sample reconstruction illustrating rippling artefacts in some of the reconstructed images. These artifacts seem to be present on images reconstructed by models that used coil sensitivity estimation and coil channel combination as part of their method. }
\label{sensitivity_issue}   
\end{centering}
\end{figure*}

All of the models submitted to the MC-MRI Reconstruction Challenge had a relatively narrow input convolutional layer (\textit{e.g.}, 64 filters), which may have resulted in the loss of relevant information. In~\cite{sriram2020grappanet}, they used 15-channel data and the first layer had 384 filters. Another advantage of models that receive all channels as input is that they seem more robust to rippling effects that can occur in the reconstructed images due to problems in coil sensitivity estimation. This finding was observed in our visual assessment (Figure~\ref{sensitivity_issue}) and consists of rippling artifacts in the reconstructions most likely due to problems in coil sensitivity estimation (ResoNNance and The Enchanted). Similar artifacts were not observed in images produced on models that do not require coil sensitivity estimation.          

In our study, we also noted variability in the ranking across metrics (Table~\ref{t1_table}). For example, The Enchanted 1.0 submission had the best VIF score, but only the fourth best SSIM and seventh highest pSNR metrics. This variability reinforces the importance of including many benchmarks that can summarise the result of multiple submissions by using a consistent set of multiple metrics. Studies that use a single image quality metric, for example, are potentially problematic if the chosen measure masks specific classes of performance issues. While imperfect, the use of a composite score based on metric rankings attempts to reduce this inherent variability by examining multiple performance measures.

\begin{figure*}[!ht]
\begin{centering}
\includegraphics[width=0.95\textwidth]{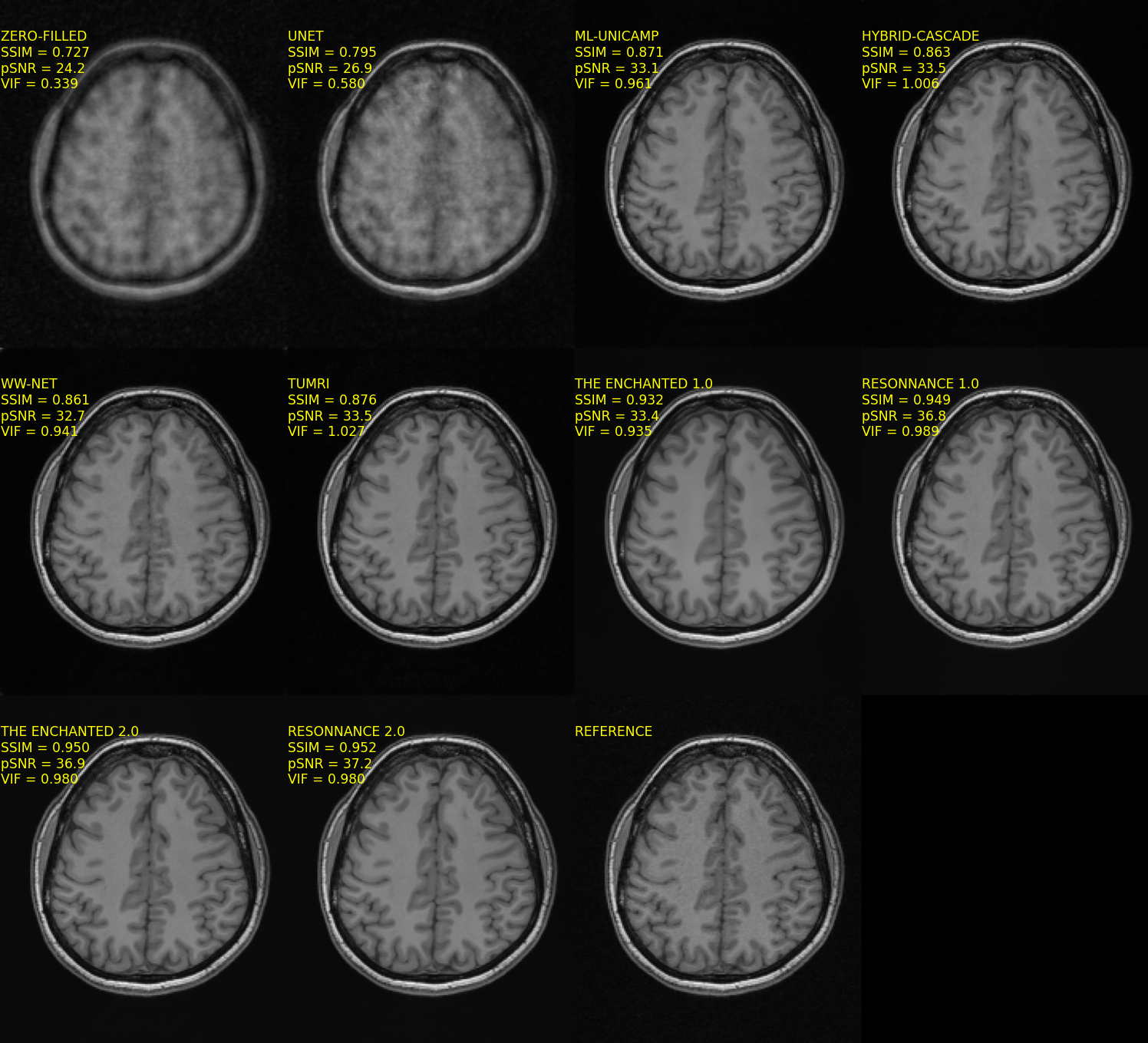}\\
\caption{Representative reconstructions of the different models submitted to Track 01 (\textit{i.e.}, 12-channel) of the challenge for $R = 5$.}
\label{sample_rec_12_channel}   
\end{centering}
\end{figure*}

\begin{figure*}[!ht]
\begin{centering}
\includegraphics[width=0.95\textwidth]{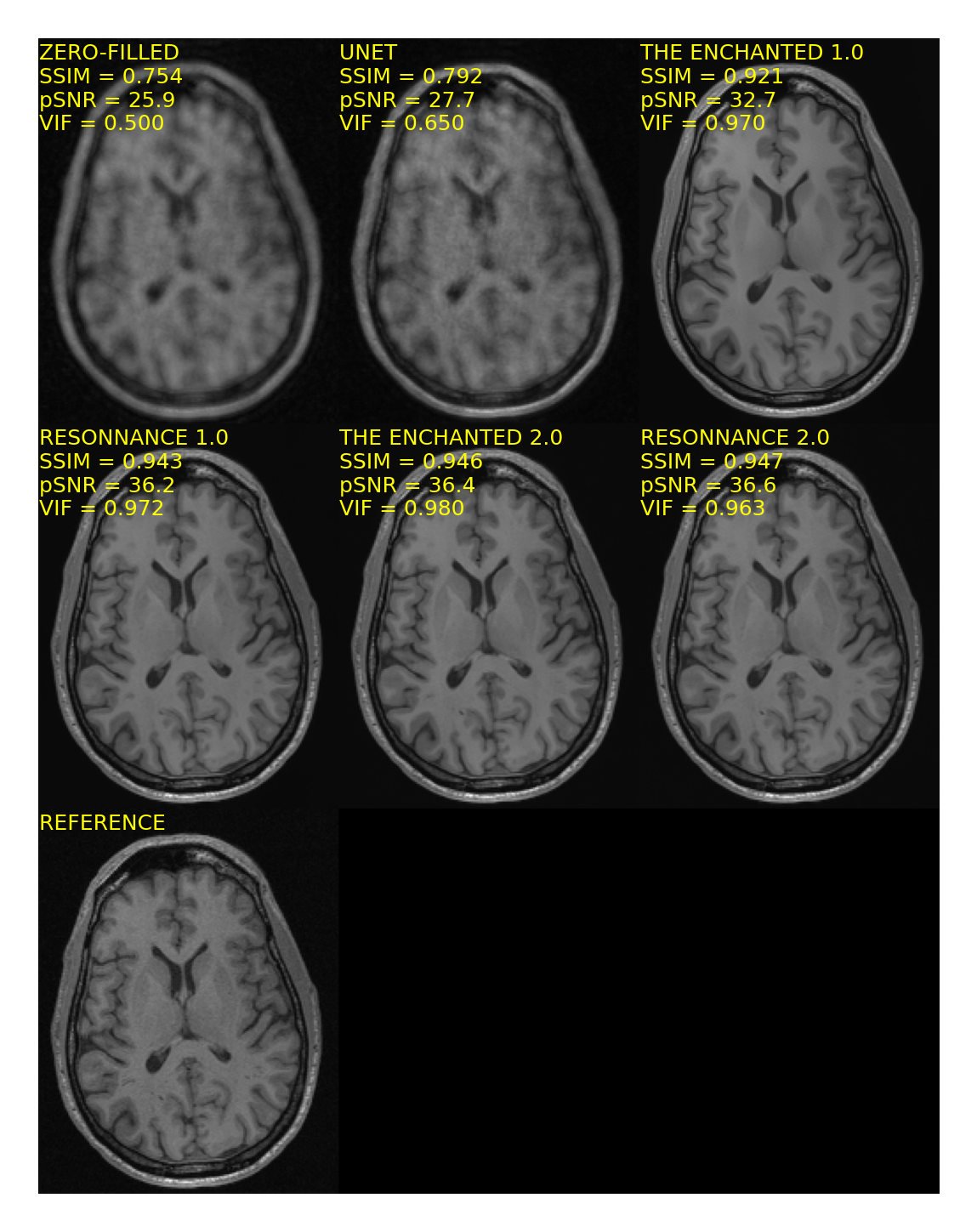}\\
\caption{Representative reconstructions of the different models submitted to Track 02 of the challenge for $R = 5$ using the 32-channel coil.}
\label{sample_rec_32_channel}   
\end{centering}
\end{figure*}

Visual inspection of the reconstructed MR images (\textit{cf.}, Figure~\ref{sample_rec_12_channel} and Figure~\ref{sample_rec_32_channel}) indicates that with some models and for some samples in the test set, the reconstructed background noise is different from the background noise in the reference images. This observation, particularly with the ResoNNance and The Enchanted teams' submissions, leads to questions on whether the evaluated quantitative metrics are best suited to determine the reconstruction quality. Given a noisy reference image, a noise-free reconstruction will potentially achieve lower pSNR, SSIM, and VIF than the same reconstruction with added noise. This finding is contrary to human visual  perception, where noise impacts the image quality negatively and is, in general, undesired. During the expert visual assessment, 23 of 50 ($46.0\%$) reconstructions were rated higher than the fully sampled reference due the fact that the brain anatomical borders in these images were preserved, but the image background noise was filtered out.  

All trainable baseline models and the model submitted by M-L UNICAMP used mean squared error as their cost function. The model submitted by TUMRI was trained using a combination of multi-scale SSIM~\cite{msssim} and VIF as their cost function. The model The Enchanted 1.0  has two components in their cost function: 1) their model was trained using mean squared error as the cost function with the target being the coil-combined complex-valued fully sampled reference and then 2) their Down-Up network~\cite{yu2019deep} received as input the absolute value of the reconstruction obtained in the previous stage and the reference was the square-root sum-of-squares fully sampled reconstruction. The Down-Up network was trained using SSIM as the loss function.
The model The Enchanted 2.0 is the only model that was pre-trained using a self-supervised learning pretext task of predicting rotations. The pretext task was trained using cross-entropy as the loss fucntion. The main task (\textit{i.e.}, reconstruction task) was trained using SSIM as the loss function.

The ResoNNance 1.0 and 2.0 models used a combination of SSIM and mean absolute error as the training loss function, which is a combination that has been shown to be effective for image restoration~\cite{zhao2016loss}. Because the background in the images is quite substantial and SSIM is a bounded metric that is computed across image patches, this observation causes models trained using SSIM as part of their loss function to try to match the background noise in their reconstructions. This observation may offer a potential explanation why the models submitted by  The Enchanted and ResoNNance teams were able to preserve the noise pattern in their reconstructions. 

For $R=5$, the top  three models: ResoNNance 2.0, The Enchanted 2.0, and ResoNNance 1.0 produced the most visually pleasing reconstructions and also had the top performing metrics.  It is important to emphasize that $R=5$ in the challenge is relative to the $85\%$ of k-space that was sampled in the slice-encode ($k_z$) direction. If we consider the equivalent full $k$-space, the acceleration factor would be $R=5.9$. Based on the Track 01 results, we would say that an acceleration between 5 and 6 might be feasible to be incorporated into a clinical setting for a single-sequence MR image reconstruction model. Further analysis of the image reconstructions by a panel of radiologists is needed to better assess clinical value before achieving a definite conclusion.

The second track of the challenge compared six different reconstruction models (Table~\ref{t1_table} and Table~\ref{t2_table_32}). The models The Enchanted 2.0 and ResoNNance 2.0  achieved the best overall results. For the 12-channel test set (Figure~\ref{sample_rec_12_channel}), the results were the same as the results they obtained in Track 01 of the challenge since the models were the same. More interesting are the results for the 32-channel test set. Though the metrics for the 32-channel test set are higher than for the 12-channel test set, by visually inspecting the quality of the reconstructed images, it is clear that 32-channel image reconstructions are of poorer quality compared to 12-channel reconstructions (Figure~\ref{sample_rec_32_channel}). Twenty eight percent of the 32-channel images assessed by the expert observer were deemed to have  poorer quality when compared to the reference compared to $4$\% of the 12-channel images rated. This fact raises concerns about generalizability of the reconstruction models across different coils. Potential approaches to mitigate this issue is to include representative data collected with different coils in the training and validation sets or employ domain adaptation techniques \cite{kouw2019review}, such as data augmentation strategies that  simulate data acquired under different coil configurations, to make the models more generalizable.

Though the generalization of learned MR image reconstruction models and their potential for transfer learning have been previously assessed~\cite{knoll2019assessment}, the results from Track 02 of our challenge indicate that there is still room for improvements. Interestingly, the model The Enchanted 2.0 is the only model that employed self-supervised learning, which seems to have had a positive impact on the model generalizability for the 32-channel test data.

\begin{figure*}
\centering 
\includegraphics[width=0.99\textwidth]{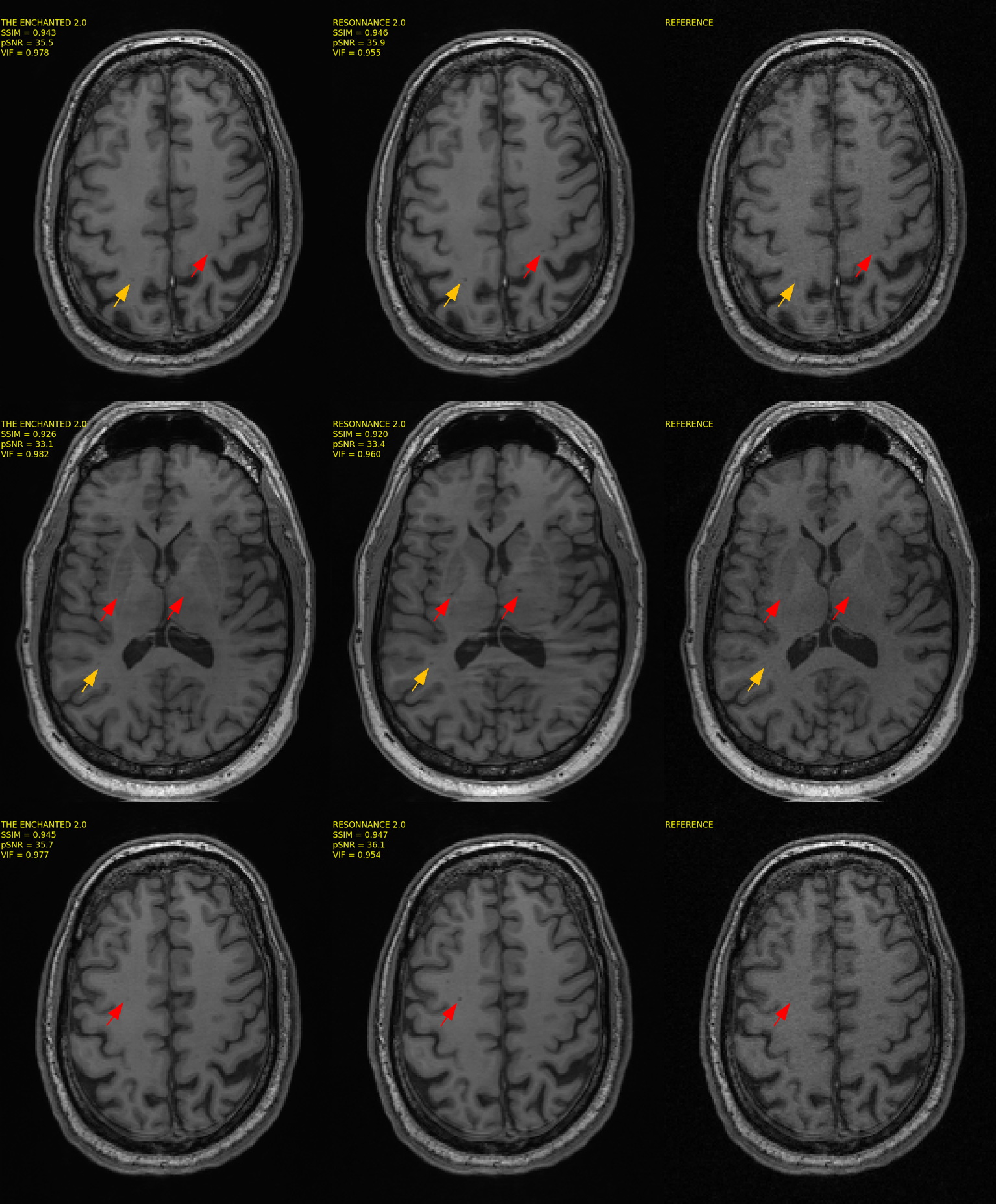}
\caption{Three sample reconstructions, one per row, for the two-top models. The Enchanted 2.0 and ResoNNance 2.0, and the reference are illustrated. The arrows in the figure indicate regions of interest that indicate deviations between the deep-learning-based reconstructions and the fully sampled reference. }
\label{bad_12}
\end{figure*}

One important finding that we noticed during visual assessment of the images is that some of the reconstructed images enhanced hypointensity regions within the brain white-matter, while in others images these hypointensities were blurred out of the image (\textit{cf.}, Figure \ref{bad_12}). In many cases, it was unclear from the fully sampled reference whether this hypointensity regions corresponded to noise in the image or if they indicated the presence of relevant structures, such as lacunes. This finding is critical and further investigation is necessary to determine its potential impact before clinical adoption of these reconstruction models.

\section{Summary}
The MC-MRI Reconstruction Challenge provided an objective benchmark for assessing brain MRI reconstruction and the generalizability of models across datasets collected with different coils using a high-resolution, 3D dataset of T1-weighted MR images. Track 01 compared ten reconstruction models and Track 02 compared six reconstruction models. The results indicated that although the quantitative metrics are higher for the test data not seen during training (\textit{i.e.}, 32-channel data), visual inspection indicated that these reconstructed images had poorer quality. This conclusion that current models do not generalize well across datasets collected using different coils indicates a promising research field in the coming years that  is very relevant for the potential  clinical adoption of deep-learning-based MR image reconstruction models. The results also indicated the difficulty of reconstructing finer details in the images, such as lacunes. The MC-MRI Reconstruction Challenge continues and the organizers of the benchmark will periodically incorporate more data, which will potentially allow to train deeper models. As a long term benefit of this challenge, we expect that the adoption of these deep-learning-based MRI reconstruction models into the clinical and research environments will be streamlined.  

\section*{Declaration of Competing Interest}
M.W.A. Caan is shareholder of Nico.lab International Ltd.

\section*{Acknowledgements}
Richard Frayne thanks the Canadian Institutes for Health Research (CIHR, FDN-143298) for supporting the Calgary Normative Study and acquiring the raw datasets. Richard Frayne and Roberto Souza thank the Natural Sciences and Engineering Research Council (NSERC, RGPIN-2021-02867, PI: Souza and XXX, PI: Frayne) also provided ongoing operating support for this project. We also acknowledge the infrastructure funding provided by the Canada Foundation of Innovation (CFI). The organizers of the challenge also acknowledge Nvidia for providing a Titan V graphics processing unit and Amazon Web Services for providing computational infra-structure that was used by some of the teams to develop their models. Dimitrios Karkalousos and Matthan Caan  were supported by the STAIRS project under the Top Consortium for Knowledge and Innovation-Public, Private Partnership (TKI-PPP) program, co-funded by the PPP Allowance made available by Health Holland, Top Sector Life Sciences \& Health. Helio Pedrini thanks the National Council for Scientific and Technological Development (CNPq \#309330/2018-1) for the research support grant. Leticia Rittner also thanks the National Council for Scientific and Technological Development (CNPq \#313598/2020-7) and São Paulo Research Foundation (FAPESP \#2019/21964-4) for the support. 

\bibliographystyle{elsarticle-num-names}
\bibliography{sample.bib}
\newpage 

\begin{table}[!h]
\caption{Summary of the Track 01 results for $R=5$. The best value for each metric and acceleration is emboldened. Mean $\pm$ standard deviation are reported. $\ast$ indicates a baseline model.}
\label{t1_table}
\centering
\resizebox{\columnwidth}{!}{%
\begin{tabular}{c|c|c|c}
\hline
\textbf{Model} &  \textbf{SSIM} &  \textbf{pSNR (dB)} &  \textbf{VIF} \\ \hline
\textbf{ResoNNance 2.0} &  $\boldsymbol{0.941 \pm 0.029}$ & $\boldsymbol{35.7 \pm 1.8}$ & $0.957 \pm 0.034$ \\ \hline
\textbf{The Enchanted 2.0} &  $0.937 \pm 0.033$ & $34.9 \pm 2.4$ & $0.973 \pm 0.036$ \\ \hline
\textbf{ResoNNance 1.0} &   $0.936 \pm 0.031$ & $35.3 \pm 1.8$ & $0.960 \pm 0.035$ \\ \hline
\textbf{The-Enchanted 1.0} &  $0.912 \pm 0.034$ & $30.3 \pm 2.8$ & $\boldsymbol{0.993 \pm 0.176}$ \\ \hline
\textbf{TUMRI} &  $0.868 \pm 0.044$ & $32.5 \pm 1.7$ & $0.989 \pm 0.045$ \\ \hline
\textbf{WW-Net$\ast$} &  $0.870 \pm 0.043$ & $32.5 \pm 1.7$ & $0.929 \pm 0.049$ \\ \hline
\textbf{Hybrid-cascade$\ast$} &  $0.860 \pm 0.044$ & $32.7 \pm 1.6$ & $0.954 \pm 0.042$ \\ \hline
\textbf{M-L UNICAMP} &  $0.868 \pm 0.044$ & $32.4 \pm 1.7$ & $0.918 \pm 0.053$ \\ \hline
\textbf{U-Net$\ast$} &  $0.779 \pm 0.039$ & $26.8 \pm 1.7$ & $0.642 \pm 0.068$ \\ \hline
\textbf{Zero-filled$\ast$} &  $0.726 \pm 0.045$ & $25.2 \pm 1.5$ & $0.518 \pm 0.066$ \\ \hline
\end{tabular}}
\end{table}

\begin{table}
\caption{Summary of the Track 02 results for $R=5$ using the 32-channel test set. The best value for each metric and acceleration is emboldened. Mean $\pm$ standard deviation are reported. $\ast$ indicates a baseline model.}
\label{t2_table_32}
\centering
\resizebox{\columnwidth}{!}{%
\begin{tabular}{c|c|c|c}
\hline
\textbf{Model} &  \textbf{SSIM} &  \textbf{pSNR (dB)} &  \textbf{VIF}\\ \hline
\textbf{ResoNNance 2.0} &  $\boldsymbol{0.961 \pm 0.027}$ & $38.3 \pm 2.2$ & $0.955 \pm 0.036$ \\ \hline
\textbf{The Enchanted 2.0} &  $0.960 \pm 0.037$ & $\boldsymbol{38.34 \pm 3.2}$ & $\boldsymbol{1.024 \pm 0.034}$ \\ \hline
\textbf{ResoNNance 1.0} &  $0.947 \pm 0.033$ & $37.7 \pm 2.9$ & $0.992 \pm 0.030$ \\ \hline
\textbf{The Enchanted 1.0} &  $0.907 \pm 0.046$ & $30.1 \pm 2.7$ & $0.834 \pm 0.236$ \\ \hline
\textbf{U-Net$\ast$} &  $0.832 \pm 0.058$ & $31.5 \pm 2.6$ & $0.804 \pm 0.045$ \\ \hline
\textbf{Zero-filled$\ast$} &  $0.780 \pm 0.041$ & $26.4 \pm 1.5$ & $0.472 \pm 0.064$ \\ \hline
\end{tabular}}
\end{table}

\end{document}